\documentclass[10pt, final, journal, letterpaper, twocolumn]{IEEEtran}
\makeatletter
\def\ps@headings{%
\def\@oddhead{\mbox{}\scriptsize\rightmark \hfil \thepage}%
\def\@evenhead{\scriptsize\thepage \hfil \leftmark\mbox{}}%
\def\@oddfoot{}%
\def\@evenfoot{}}
\makeatother 

\newcommand{\bs}{\boldsymbol}

\IEEEoverridecommandlockouts

\usepackage{cite}
\usepackage{amsmath,amsthm,amssymb,amsfonts}
\usepackage{algorithm}
\usepackage{algorithmic}
\usepackage{graphicx}
\usepackage{color}
\usepackage[normalsize]{subfigure}

\newtheorem{remark}{Remark}

\newtheorem{lemma}{Lemma}

\begin{document}

\title{Artificial Noise Aided Secrecy Information and Power Transfer in OFDMA Systems}

\author{\IEEEauthorblockN{Meng Zhang,~\IEEEmembership{Student Member,~IEEE},
 Yuan Liu,~\IEEEmembership{Member,~IEEE},
 and Rui Zhang,~\IEEEmembership{Senior Member,~IEEE}}

%
%
%
}

\maketitle

\vspace{-1.5cm}
\begin{abstract}
 In this paper, we study simultaneous wireless information and power transfer (SWIPT) in orthogonal frequency division multiple access (OFDMA) systems with the coexistence of information receivers (IRs) and energy receivers (ERs). The IRs are served with best-effort secrecy data and the ERs harvest energy with minimum required harvested power. To enhance the physical layer security for IRs and yet satisfy energy harvesting requirements for ERs, we propose a new frequency-domain artificial noise (AN) aided transmission strategy. With the new strategy, we study the optimal resource allocation for the weighted sum secrecy rate maximization for IRs by power and subcarrier allocation at the transmitter. The studied problem is shown to be a mixed integer programming  problem and thus  non-convex, while we propose an efficient algorithm for solving it based on the Lagrange duality method. To further reduce the computational complexity, we also propose a suboptimal algorithm of lower complexity. The simulation results illustrate the effectiveness of proposed algorithms as compared against other heuristic schemes.
\end{abstract}

\begin{keywords}
Physical layer security, simultaneous wireless information and power transfer (SWIPT), artificial noise (AN), orthogonal frequency-division multiple access (OFDMA), resource allocation.
\end{keywords}

\section{Introduction}

\IEEEPARstart{O}RTHOGONAL frequency division multiple access (OFDMA) has many advantages such as flexibility in resource allocation and robustness against multipath channel fading, and therefore has become a well established multiple-access technique for multiuser wireless communications systems.

Recently, simultaneous wireless information and power transfer (SWIPT) provides an appealing solution to prolong the operation time of energy-limited wireless nodes \cite{ZhangHo,Zhou,Liuz,LiuPS,NgOFDM,ZhouZhang,DerrickIC13,ER10}. SWIPT systems enable the users to harvest energy and decode information from the same received signal, thus making most efficient use of the wireless spectrum for both information and energy transfer. SWIPT has drawn a great amount of research interests. For instance,
two practical schemes for SWIPT, namely power splitting (PS) and time switching (TS), were proposed in \cite{ZhangHo} and \cite{Zhou}. With TS applied at each receiver, the received signal is either processed for energy harvesting or for information decoding. When PS is used at the receiver, the signal is split into two streams, for information decoding and energy harvesting, respectively. The authors in \cite{ZhangHo} and \cite{Zhou} also investigated the achievable rate-energy tradeoffs for a multiple-input multiple-output (MIMO) SWIPT system and a single-input single-output (SISO) SWIPT system, respectively. SWIPT systems in fading channels were studied by dynamic time switching (DTS) and dynamic power splitting (DPS) in \cite{Liuz} and \cite{LiuPS}, respectively.
    \begin{figure}[t]
\begin{centering}
\includegraphics[scale=0.8]{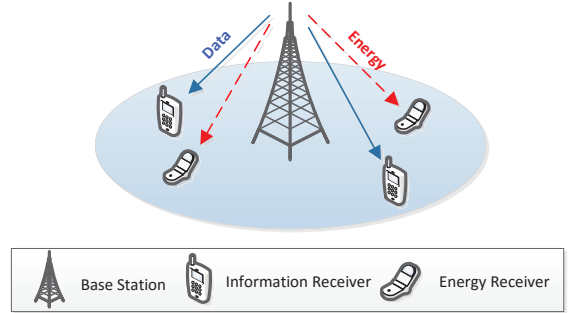}
\vspace{-0.1cm}
 \caption{System model of OFDMA-based SWIPT, where each receiver is a potential eavesdropper to other receivers. }\label{fig:sys1}
\end{centering}
\vspace{-0.3cm}
\end{figure}

On the other hand, due to the increasing importance of information security, substantial research efforts have been dedicated to information-theoretic physical layer security \cite{Leung,Wang,Jorswieck,Renna,Goel,Qin13,McKay2010,MISORobust}, as a complementary solution to the traditional cryptography based encryption applied in the upper layers. The authors in \cite{Wang} considered physical layer security in an OFDMA system, with the goal of maximizing the sum rate of best-effort information users subject to the individual secrecy rate requirements of secure users. In \cite{Renna}, the orthogonal frequency division multiplexing (OFDM) based wiretap channel was considered and the achievable secrecy rate with Gaussian inputs was studied. Artificial noise (AN) is a well-known approach for improving physical layer security by degrading eavesdroppers' channel condition \cite{Qin13,Goel}. In \cite{Goel}, in order to assist secrecy information transmission, AN is transmitted into the null space of the channels of legitimate users to interfere with the eavesdroppers. In \cite{Qin13}, the authors proposed a time-domain AN design by exploiting temporal degrees of freedom from the cyclic prefix in OFDM modulated signals, even with a single antenna at the transmitter. In \cite{MISORobust}, the authors studied robust transmission schemes for the multiple-input single-output (MISO) wiretap channels.

\begin{figure*}[t!]
\begin{centering}
\includegraphics[scale=1]{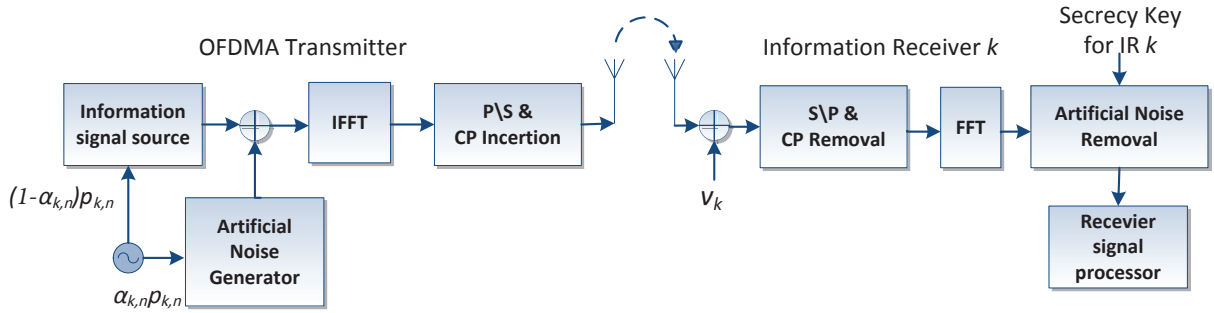}
\vspace{-0.1cm}
 \caption{Block diagram of an OFDMA transmitter and receiver with AN generation and removal procedure. }\label{fig:sys2}
\end{centering}
\vspace{-0.3cm}
\end{figure*}

A handful of works have been investigated the secure issues in SWIPT systems\cite{LiuGC13,NgGC13,Rui14,TIFS}. Since the energy receivers (ERs) need to be deployed much closer to the access points than the information receivers (IRs) due to their much higher received power requirement\cite{ER10}, they are inevitably capable of eavesdropping the messages to the IRs. Moreover, AN also plays a role of energy signal, i.e., besides interfering with the eavesdroppers to facilitate secure communication, AN is a new source for wireless power transfer as well.
In \cite{LiuGC13,NgGC13}, the authors studied the secrecy communication in SWIPT by properly designing the beamforming vectors at the multi-antenna transmitter.
Secrecy communication in SWIPT over fading channels was also studied in \cite{Rui14}. In \cite{TIFS}, the authors studied the secure OFDMA-based systems with a power splitter applied at each user terminal to coordinate the secure transmission and energy harvesting. However, AN aided OFDMA-based SWIPT systems with secrecy constraints have not yet been investigated in the literature. In a secure OFDMA system without AN, only the user with the largest channel gain  over each subcarrier (SC) can receive secure information \cite{Wang}. Thus, the new method of using AN not only achieves the secrecy information and wireless power transfer at the same time, but also leads to new  resource allocation solutions different from the conventional secure OFDMA system without AN.

Motivated by the aforementioned reasons, in this paper, we study the optimal resource allocation in the AN aided secure OFDMA systems with SWIPT as shown in Fig. \ref{fig:sys1}, where two types of receivers are assumed, i.e., IRs and ERs. Our goal is to maximize the weighted sum secrecy rate of the IRs subject to minimum harvested power requirements of individual ERs. We propose a new frequency-domain AN method in OFDMA-based SWIPT to facilitate both secrecy information transmission and energy transfer to IRs and ERs, respectively. 
Specifically, as shown in Fig. \ref{fig:sys2}, independent AN is added over each SC at the transmitter and only the desired IR is able to cancel it using the corresponding key before decoding the information\footnote{Note that the key-assisted approach is normally exclusively used for cryptography, while physical-layer methods are traditionally adopted when the shared keys are not available. However, some recent works (e.g. \cite{CC,Tan}) have considered applying  physical-layer security to enhance cryptographic secrecy, showing the potentials to benefit from both types of secrecy approaches. Hence, under such multi-layer security framework, it is also possible to jointly consider the key-assisted physical-layer security and cryptography design, which is left for our future work.}. The formulated problem is a mixed integer programming problem and thus non-convex. We propose an efficient algorithm based on the Lagrange duality method, which solves the problem asymptotically optimally when the number of SCs becomes large. Moreover, a suboptimal algorithm is also proposed to tradeoff complexity and performance.


The remainder of this paper is organized as follows. In Section II, we introduce the system model of the OFDMA-based SWIPT with secrecy constraints, and present the problem formulation. The problem is solved by the Lagrange duality method in Section III. In Section IV, we propose the suboptimal algorithm of lower complexity. In Section V, we provide the numerical results on the performance of proposed algorithms. Finally, we conclude the paper in Section VI.

\section{System Model and Problem Formulation}
%

    We consider a downlink OFDMA-based SWIPT system with secrecy constraints as shown in Fig. \ref{fig:sys1}. The system consists of one base station (BS) with a single antenna, $K$ single-antenna receivers and $N$ SCs. The set of receivers is denoted by $\mathcal{K}=\{1,...,K\}$, among which $K_1$ receivers are IRs given by the set $\mathcal{K}_1$ and the rest $K_2$ receivers are ERs given by the set $\mathcal{K}_2$, i.e., $\mathcal{K}_1\cup \mathcal{K}_2 = \mathcal{K}$. Note that the receivers (both IRs and ERs) are considered to be \textit{separated} and can only decode information or harvest energy at a time, unlike the \textit{co-located} receivers considered in \cite{NgOFDM,ZhouZhang}. The set of SCs is denoted as $\mathcal{N}=\{1,...,N\}$. Furthermore, we assume that for each IR, all other receivers (IRs and ERs) are potential eavesdroppers, similar to the case considered in \cite{Wang}. The BS is assumed to know the channel state information (CSI) of all receivers. This is practically valid since the IRs and ERs need to help the BS in obtaining their individual CSI for receiving required information and energy, respectively. We assume that the OFDM symbols are time slotted so that the length of each time slot is comparable to the channel coherence time, i.e., the channel impulse response can be treated as time invariant during each time slot. As a result, the BS can accurately estimate CSI of all receivers on all SCs.

    We propose a frequency-domain AN generation and removal method for OFDMA-based SWIPT, similar to that in \cite{Rui14} over the time domain. The scheme is illustrated in Fig. \ref{fig:sys2} and described as follows. A large ensemble of sequences used to generate Gaussian distributed AN are pre-stored at the BS\footnote{Note that in the literature, the AN is usually assumed to follow Gaussian distribution (e.g. \cite{Goel,Qin13,McKay2010}).}, and the indices of the sequences are regarded as the keys. After SC allocation to IRs, the BS first randomly picks $N$ sequences (each corresponds to one SC) from the ensemble and transmits each of their indices (keys) to the IR to whom the corresponding SC is assigned. As the random sequence (or AN) is only known to the intended IR but unknown to all the other receivers, any potential eavesdropper cannot have access to the random sequence used at each SC. Moreover, in order to prevent the eavesdropper from decoding the random sequence by long-term observation of the signal, the BS randomly picks new random sequences and transmits the corresponding keys in a secret manner to the desired IRs from time to time, using e.g. the method proposed in \cite{Koo2000} by exploiting the channel independence and reciprocity. Specifically, the IR sends a pilot signal to the BS, and then the BS sends a random key and modulates it over the phase of the transmitted signal with the received channel phase pre-compensated. In this way, the intended IR is able to decode the key while the channel phases between the BS and other receivers are different from that between the BS and the intended IR. Thus, the key can be confidentially transmitted without being eavesdropped by other receivers.

    The transmit signal comprises the transmitted data symbol $s_{k,n}$ from the BS to IR $k$ on SC $n$  and the AN bearing signal $z_{k,n}$ for IR $k$, $k \in \mathcal{K}_1$ and $n \in \mathcal{N}$. It is assumed that $s_{k,n}$ and $z_{k,n}$ are independent circularly symmetric complex Gaussian (CSCG) random variables with zero mean and unit variance, denoted by $s_{k,n}\sim\mathcal{CN}(0,1)$ and $z_{k,n}\sim\mathcal{CN}(0,1)$, which are also independent over $n$.

    The transmitted signal to IR $k$ at SC $n$ is given by
    \begin{equation}
    X_{k,n}=\sqrt{(1-\alpha_{k,n})p_{k,n}}s_{k,n}+\sqrt{\alpha_{k,n}p_{k,n}}z_{k,n},
    \end{equation}
     where $p_{k,n}\geq 0$ is the total power at SC $n$ and $0\leq\alpha_{k,n}\leq1$ is the transmit power splitting ratio at the BS-side to generate AN to be added at SC $n$.

     Let $h_{k,n}$ denote the complex channel coefficient from the BS to receiver $k$ at SC $n$, and $\beta_{k,n}$ denote the eavesdropper's complex channel coefficient. Here, we let $|\beta_{k,n}|^2=\max_{k' \in \mathcal{K}, k'\neq k}|h_{k',n}|^2$, indicating that the considered eavesdropper of receiver $k$ is the receiver of the largest channel gain among all the other receivers on SC $n$.
     The downlink received signal at IR $k$ on SC $n$ and that at a potential eavesdropper who is wiretapping IR $k$ over SC $n$ are respectively given by
     \begin{align}
    &Y_{k,n}=h_{k,n}X_{k,n}+v_k ,\\
    &E_{k,n}=\beta_{k,n}X_{k,n}+e_k,
    \end{align}
    where the noise $v_k$ and $e_k$ are assumed to be independent and identically distributed (i.i.d.) as  $\mathcal{CN}(0,\sigma^2)$.

    With the aforementioned scheme, the AN can be canceled at the desired IR at each SC but not possibly at any of the potential eavesdroppers. From (1)-(3), the received signals at IR $k$ after AN cancelation and the ``best" eavesdropper on SC $n$ are further expressed as
     \begin{align}
    &Y_{k,n}=h_{k,n}\sqrt{(1-\alpha_{k,n})p_{k,n}}s_{k,n}+v_k ,\\
    &E_{k,n}=\beta_{k,n}\sqrt{(1-\alpha_{k,n})p_{k,n}}s_{k,n}+\beta_{k,n}\sqrt{\alpha_{k,n}p_{k,n}}z_{k,n}+e_k.
    \end{align}
Here we can write the achievable information rate of IR $k$ on SC $n$, which is given by
      \begin{align}
r_{k,n}=\log_2 \left(1+\frac{(1-\alpha_{k,n})|h_{k,n}|^2p_{k,n}}{\sigma^2}\right).
      \end{align}
The decodable information rate of the ``best" eavesdropper on SC $n$ is given by
 \begin{align}
r_{k,n}^e=\log_2 \left(1+\frac{(1-\alpha_{k,n})|\beta_{k,n}|^2 p_{k,n}}{\sigma^2+\alpha_{k,n}|\beta_{k,n}|^2 p_{k,n}}\right).
 \end{align}
The achievable secrecy rate for IR $k$ on SC $n$ is thus given by \cite{Shannon}
\begin{align}\label{r}
R_{k,n}^s=&[r_{k,n}-r_{k,n}^e]^+\nonumber\\
=&\left[\log_2 \left(1+\frac{(1-\alpha_{k,n})|h_{k,n}|^2p_{k,n}}{\sigma^2}\right)\right.\nonumber\\
&\left.-\log_2 \left(1+\frac{(1-\alpha_{k,n})|\beta_{k,n}|^2p_{k,n}}{\alpha_{k,n}|\beta_{k,n}|^2p_{k,n}+\sigma^2}\right)\right]^+,
\end{align}
for all $k \in \mathcal{K}_1$ and $n \in \mathcal{N}$, where $[\cdot]^+\triangleq \max(0,\cdot)$.

\begin{lemma}
$R_{k,n}^s$ in \eqref{r} can be further expressed as
\begin{equation}
R_{k,n}^s=\begin{cases}~0, &~{\rm if}~ 0\leq p_{k,n}\leq [\mathcal{X}_{k,n}(\alpha_{k,n})]^+,\\
         ~r_{k,n}-r_{k,n}^e\geq0, &~{\rm if}~p_{k,n}>[\mathcal{X}_{k,n}(\alpha_{k,n})]^+, \end{cases}
\end{equation}
where

\begin{equation}
\mathcal{X}_{k,n}(\alpha_{k,n}) \triangleq \begin{cases} \frac{\sigma^2}{\alpha_{k,n}}\left(\frac{1}{|h_{k,n}|^2}-\frac{1}{|\beta_{k,n}|^2}\right) &~{\rm if}~ \alpha_{k,n}\neq0\\
{\rm sgn}\left(|\beta_{k,n}|^2-|h_{k,n}|^2\right)\times\infty&~{\rm if}~ \alpha_{k,n}=0~~ \end{cases},
\end{equation}
and ${\rm sgn}(x)=|x|/x$ if $x\neq0$ and ${\rm sgn}(x)=1$ if $x=0$.

\end{lemma}
\begin{IEEEproof}
Please refer to Appendix A.
\end{IEEEproof}

\begin{remark}
 Note that the traditional AN scheme (without AN cancelation, e.g. \cite{Goel,MISORobust}) is
ineffective for the considered SISO systems, i.e., without cancelling AN in the intended IRs, AN cannot achieve a higher secrecy rate compared to the transmission without AN. The details can be found in Appendix B.
\end{remark}

The weighted sum (secrecy) rate of all $K_1$ IRs is given by
\begin{equation}\label{sr}
R_{\rm sum}^s=\sum_{k \in \mathcal{K}_1} w_k \sum_{n \in \mathcal{N}} x_{k,n} R_{k,n}^s,
\end{equation}
where $w_k$ is the positive weight of IR $k$ and $x_{k,n}$ is the binary SC allocation variable with $x_{k,n}=1$ representing SC $n$ is allocated to IR $k$ and $x_{k,n}=0$ otherwise. Note that in the considered system, the ERs can harvest energy from all SCs while the IRs need orthogonal SC assignment for avoiding mutual interference. In addition, if the power allocated on SC $n$ is given by $p_n$, then ER $l$ can harvest $\zeta_l p_{n}|h_{l,n}|^2$ on SC $n$ regardless of which receiver it is allocated to. Notice that if $p_{k,n}>0$ and $\alpha_{k,n}=1$ for any SC $n$, then this SC is used only for energy transfer, i.e., there is no information sent over the SC. As a result, we only need to focus on the cases that SCs are allocated to IRs without loss of generality.

Thus, the harvested power at each ER $l \in \mathcal{K}_2$ is given by
\begin{align}
Q_{l}=\zeta_l \sum_{n \in \mathcal{N}}\left(\sum_{k \in \mathcal{K}_1} x_{k,n} p_{k,n}\right)|h_{l,n}|^2, \label{eqn:EH}
\end{align}
where $0<\zeta_{l}<1$ denotes the energy harvesting efficiency.

   \begin{figure}[t]
\begin{centering}
\includegraphics[scale=0.45]{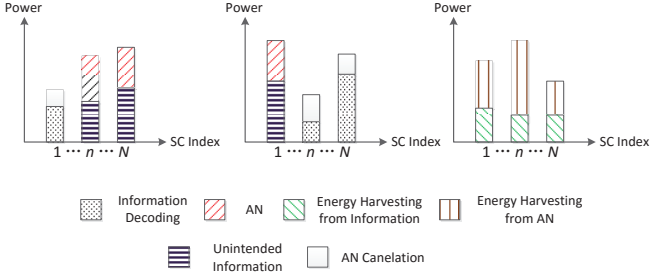}
\vspace{-0.1cm}
 \caption{An example of power utilization for an OFDMA-based SWIPT system of two IRs and one ER. }\label{gaga}
\end{centering}
\vspace{-0.3cm}
\end{figure}

An example of the energy utilization at receivers in an OFDMA-based SWIPT system with secrecy constraints is shown in Fig. \ref{gaga}, with $K_1=2$ and $K_2=1$. As it is shown, the AN does not interfere with the intended receiver but all other receivers. In addition, the ER is able to harvest energy from both information signal and AN signal.

%
%

 Our goal is to maximize the weighted sum rate of the IRs by optimizing transmit power and SC allocation as well as transmit power splitting ratio at each SC, subject to the harvested power constraints of all ERs.
 The problem can be mathematically formulated as
 \begin{subequations}
\begin{align}
 &\max_{\bs P,\bs X,\bs \alpha} R_{\rm sum}^s \label{eqn:pbp}\\
{\rm s.t.}~~~
& Q_{l} \geq \bar{Q}_l, \forall l \in \mathcal{K}_2, \label{con:r}\\
&\sum_{k \in \mathcal{K}_1}\sum_{n \in \mathcal{N}} p_{k,n}x_{k,n} \leq P_{\rm max} \label{con:p1}\\
&0\leq p_{k,n} \leq P_{\rm peak}, \forall n \in \mathcal {N},k \in \mathcal{K}_1 \label{con:p2}\\
&0\leq \alpha_{k,n}\leq 1, \forall n \in \mathcal {N},k \in \mathcal{K}_1 \label{con:a}\\
&x_{k,n} \in \{0,1\}, \forall n \in \mathcal {N},k \in \mathcal{K}_1 \label{con:x1}\\
&\sum_{k \in \mathcal{K}_1} x_{k,n} \leq 1, \forall n \in \mathcal {N} ,\label{con:x2}
\end{align}
\end{subequations}
where $\bs P \triangleq \{p_{k,n}\}$ denotes the power allocation over SCs, $\bs X \triangleq \{x_{k,n}\}$ denotes the SC allocation for IRs, and $\bs \alpha \triangleq \{\alpha_{k,n}\}$ denotes the transmit power splitting over SCs.
In \eqref{con:r}, $\bar{Q}_l$ denotes the harvested power constraint for ER $l\in \mathcal{K}_2$. In \eqref{con:p1} and \eqref{con:p2}, $P_{\rm max}$ and $P_{\rm peak}$ represent the total power constraint over all SCs and the peak power constraint over each SC, respectively. Finally, \eqref{con:x1} and \eqref{con:x2} constrain that any SC can only be assigned to at most one IR.

\section{Optimal Solution}

Problem \eqref{eqn:pbp} is a mixed integer programming and thus is NP-hard and non-convex. As shown in \cite{noncon,resourceOFDMA}, the duality gap becomes zero in OFDM-based resource allocation problems including problem \eqref{eqn:pbp} as the number of SCs goes to infinity due to the so-called time-sharing condition.
 This implies that problem \eqref{eqn:pbp} can be solved by the Lagrange duality method asymptotically optimally.

 First, the Lagrangian of problem \eqref{eqn:pbp} is given by
\begin{align}
&\mathcal{L}\left(\bs P,\bs \alpha, \bs X, \bs \lambda, \gamma \right)\nonumber\\
=& \sum_{k \in \mathcal{K}_1} w_k \sum_{n \in \mathcal{N}} x_{k,n} R_{k,n}^s- \gamma \left(\sum_{k \in \mathcal{K}_1}\sum_{n \in \mathcal{N}} x_{k,n}p_{k,n} - P_{\rm max}\right)\nonumber\\
&+\sum_{l \in \mathcal{K}_2}\lambda_l(Q_{l}-\bar{Q}_l)\nonumber\\
=&\sum_{k \in \mathcal{K}_1} w_k\sum_{n \in \mathcal{N}} x_{k,n} R_{k,n}^s-\gamma\sum_{k \in \mathcal{K}_1}\sum_{n \in \mathcal{N}} x_{k,n}p_{k,n} \nonumber\\
&+\sum_{n \in \mathcal{N}}\left(\sum_{k \in \mathcal{K}_1} x_{k,n} p_{k,n}\right)\sum_{l \in \mathcal{K}_2}\lambda_l\zeta_l |h_{l,n}|^2\nonumber\\
&-\sum_{l \in \mathcal{K}_2}\lambda_l \bar{Q}_l + \gamma P_{\rm max}\label{eqn:lf},
\end{align}
where $\bs \lambda=[\lambda_1,\lambda_2,...,\lambda_{K_2}]$ and $\gamma$ are the Lagrange multipliers (dual variables) corresponding to the minimum required harvested power constraints and the total transmit power constraint, respectively.

We then define $\mathcal{P}$ for given $\bs X$ as the set of all possible power allocations of $\bs P$ that satisfy $0\leq p_{k,n}\leq P_{\rm peak}$ for $x_{k,n}=1$ and $p_{k,n}=0$ when $x_{k,n}=0$, $\mathcal{S}$ as the set of all possible $\bs X$ that satisfy constraints \eqref{con:x1} and \eqref{con:x2}, and $\mathcal{A}$ as the set of all feasible $\bs \alpha$ that satisfy \eqref{con:a}. Then, we can obtain the dual function for problem \eqref{eqn:pbp} as
\begin{align}\label{eqn:dual}
g(\bs \lambda, \gamma)= \max_{\bs P \in \mathcal{P}(\bs X), \bs \alpha \in \mathcal{A}, \bs X \in \mathcal{S}}\mathcal{L} \left(\bs P,\bs \alpha, \bs X, \bs \lambda, \gamma\right).
\end{align}

The dual problem is then given by
\begin{align}\label{eqn:dp}
\min_{\bs \lambda\succeq0, \gamma\geq0}g(\bs \lambda, \gamma).
\end{align}

From \eqref{eqn:lf}, we can observe that the maximization in \eqref{eqn:dp} can be decomposed into $N$ independent subproblems. Accordingly, we can rewrite the Lagrangian as
\begin{align}
\mathcal{L}\left(\bs P,\bs \alpha, \bs X, \bs \lambda, \gamma \right)=&\sum_{n \in \mathcal{N}}\mathcal{L}_n\left(\bs P_n,\bs \alpha_n, \bs X_n\right) \nonumber\\ &-\sum_{l \in \mathcal{K}_2}\lambda_l \bar{Q}_l+ \gamma P_{\rm max},
\end{align}
where
\begin{align}
&\mathcal{L}_n
\left(\bs P_n,\bs \alpha_n, \bs X_n\right)\nonumber\\
\triangleq &\sum_{k \in \mathcal{K}_1}x_{k,n}\left\{w_k R_{k,n}^s - \gamma p_{k,n}
+p_{k,n} \left( \sum_{l \in \mathcal{K}_2}\lambda_l\zeta_l |h_{l,n}|^2\right)\right\}. \label{subl}
\end{align}
Since $x_{k,n}\in\{0,1\}$ and $\sum_{k\in\mathcal{K}_1}x_{k,n}=1$, there exists a $k^*\in\mathcal{K}_1$ such that
   \begin{eqnarray}
     x_{k,n}^*=\begin{cases} ~1, ~~~{\rm if}~k=k^*, \\
     ~0, ~~~{\rm otherwise}\end{cases}, \forall n\in \mathcal{N}, \label{eqn:opx1}
   \end{eqnarray}
is the optimal solution to maximize $\mathcal{L}$.

Hence, with given $\bs \lambda$ and $\gamma$, the maximization of $\mathcal{L}$ can be attained by selecting
\begin{equation}
k^*=\arg\max_{k \in \mathcal{K}_1}\left\{w_k R_{k,n}^s
+p_{k,n} \left( \sum_{l \in \mathcal{K}_2}\lambda_l\zeta_l |h_{l,n}|^2- \gamma\right)\right\}
\end{equation}
for each SC $n$, and the optimal $(p_{k,n}^*,\alpha_{k,n}^*)$ can be solved by assuming $k=k^*$ and then solving the following subproblem for each SC $n$,
\begin{align}
\max_{\bs P_n \in \mathcal{P}(\bs X), \bs \alpha_n \in \mathcal{A}} &\mathcal{L}'_n
\left(\bs P_n,\bs \alpha_n\right)\nonumber\\
\triangleq & w_k R_{k,n}^s
+p_{k,n} \left( \sum_{l \in \mathcal{K}_2}\lambda_l\zeta_l |h_{l,n}|^2- \gamma\right).\label{subq}
\end{align}

\subsection{Joint Optimization of Power Allocation and Transmit Power Splitting Ratio}
We cannot directly express the partial derivative of $R_{k,n}^s$ in \eqref{r} with respect to $p_{k,n}$ or $\alpha_{k,n}$. However, as we have discussed in Lemma 1, $R_{k,n}^s=0$ when $0\leq p_{k,n}\leq [\mathcal{X}_{k,n}(\alpha_{k,n})]^+ $ and $R_{k,n}^s>0$ when $p_{k,n}>[\mathcal{X}_{k,n}(\alpha_{k,n})]^+$. In each region, $R_{k,n}^s$ is differentiable with respect to $p_{k,n}$ or $\alpha_{k,n}$. Hence, we first find the set of all feasible candidates for $(p_{k,n}^*, \alpha_{k,n}^*)$ in all regions. Then, we select $(p_{k,n}^*, \alpha_{k,n}^*)$ as the one achieving the largest value of $\mathcal{L}'_n$ in \eqref{subq}.  

\subsubsection{Region I ($p_{k,n}>[\mathcal{X}_{k,n}(\alpha_{k,n})]^+$)}
\begin{lemma}
The optimal $\alpha_{k,n}$ with given $p_{k,n}$ for problem \eqref{subq} is given by
\begin{align}
\alpha_{k,n}^*(p_{k,n})=\left[\frac{1}{2}+\frac{\sigma^2}{2p_{k,n}}\left(\frac{1}{|h_{k,n}|^2}-\frac{1}{|\beta_{k,n}|^2}\right)\right]^+<1\label{eqn:opta1},
\end{align}
for all $k \in \mathcal{K}_1$ and $n \in \mathcal{N}$.

\end{lemma}

\begin{IEEEproof}
Please refer to Appendix C.
\end{IEEEproof}



On the other hand, by deriving the partial derivative of $\mathcal{L}'_n$ with respect to $p_{k,n}$ and equating it to zero, we have
\begin{align}
a_1p_{k,n}^3+b_1p_{k,n}^2+c_1p_{k,n}+d_1=0\label{eqn:optp1},
\end{align}
where
\begin{align}
a_1=&\ln 2|h_{k,n}|^2(\alpha_{k,n}^2-\alpha_{k,n})|\beta_{k,n}|^4\Omega_{n},\\
b_1=&(\alpha_{k,n}^2-\alpha_{k,n})|\beta_{k,n}|^4|h_{k,n}|^2w_k\nonumber\\
&+\ln2|\beta_{k,n}|^2\sigma^2\left[(\alpha_{k,n}^2-1)|h_{k,n}|^2-|\beta_{k,n}|^2\alpha_{k,n}\right] \Omega_{n},\\
c_1=&\ln2(\alpha_{k,n}-1)(|h_{k,n}|^2-|\beta_{k,n}|^2)\sigma^4\Omega_{n}\nonumber\\
&+2(\alpha_{k,n}^2-\alpha_{k,n})|\beta_{k,n}|^2|h_{k,n}|^2w_k\sigma^2,\\
d_1=& (\alpha_{k,n}-1)(|h_{k,n}|^2-|\beta_{k,n}|^2)w_k\sigma^4-\ln2 \sigma^6\Omega_{n},\\
\Omega_{n}=&-\gamma+\sum_{l \in \mathcal{K}_2}\lambda_l\zeta_l |h_{k,n}|^2.\label{eqn:Omega}
\end{align}
%
%
%
 We first define $\Phi_1(\alpha_{k,n})$ as the set of all non-negative real roots to \eqref{eqn:optp1} that satisfy $[\mathcal{X}_{k,n}(\alpha_{k,n})]^+< p_{k,n}\leq P_{{\rm peak}}$ with given $\alpha_{k,n}$.
 Then, we define another set $\Psi_1(\alpha_{k,n})$ as follows:
 \begin{align}
 \Psi_1(\alpha_{k,n})\triangleq\{(p_{k,n},\alpha_{k,n})|p_{k,n} \in \Phi_1(\alpha_{k,n})\}.
 \end{align}

To find feasible candidates for ($p_{k,n}^*$,$\alpha_{k,n}^*$), we consider the following two subregions.
\begin{itemize}
  \item   For \textit{subregion i}, we remove the $[\cdot]^+$ operator of $\alpha_{k,n}^*(p_{k,n})$ in \eqref{eqn:opta1} and assume that $p_{k,n}\geq\left(\frac{1}{|\beta_{k,n}|^2}-\frac{1}{|h_{k,n}|^2}\right)\sigma^2$. Substituting it into \eqref{eqn:optp1} to eliminate $\alpha_{k,n}$, we have
\begin{align}
a_2p_{k,n}^2+b_2p_{k,n}+c_2=0,\label{eqn:optp2}
\end{align}
where
\begin{align}
a_2=&\ln2 |\beta_{k,n}|^4|h_{k,n}|^2 \Omega_{n},\\
b_2=&w_k|\beta_{k,n}|^4 |h_{k,n}|^2+ \ln 2 \Omega_{n}|\beta_{k,n}|^2\sigma^2,\\
c_2=&\sigma^2\left\{|\beta_{k,n}|^2|h_{k,n}|^2w_k(1-|\beta_{k,n}|^2)\right.\nonumber\\
&\left.+\ln2 \Omega_{n}^2(|\beta_{k,n}|^2+|h_{k,n}|^2)\right\}.
\end{align}
Similarly, we define $\Phi_2$ as the set of all non-negative real roots to \eqref{eqn:optp2} that satisfy $[\mathcal{X}_{k,n}(\alpha_{k,n})]^+< p_{k,n}\leq P_{{\rm peak}}$. We further define $\Psi_2$ as the set of all feasible candidates for $(p_{k,n}^*,\alpha_{k,n}^*)$ in \textit{subregion i} as follows:
 \begin{align}
 \Psi_2\triangleq&\{(p_{k,n},\alpha_{k,n})|p_{k,n} \in \Phi_2, \alpha_{k,n}=\alpha_{k,n}^*(p_{k,n})\}\nonumber\\
 &\cup{\left(P_{\rm peak},\alpha_{k,n}^*(P_{\rm peak})\right)},
 \end{align}
where $\alpha_{k,n}^*(p_{k,n})$ is obtained in \eqref{eqn:opta1}.
  \item   For \textit{subregion ii}, $\alpha_{k,n}^*(p_{k,n})=0$ and $p_{k,n}<\left(\frac{1}{|\beta_{k,n}|^2}-\frac{1}{|h_{k,n}|^2}\right)\sigma^2$ (which can be true only when $|h_{k,n}|^2>|\beta_{k,n}|^2$). The set of all feasible candidates for $(p_{k,n}^*,\alpha_{k,n}^*)$ in this case is given by $\Psi_1(\alpha_{k,n}=0)$ obtained via \eqref{eqn:optp1}.
\end{itemize}

    \begin{figure*}[t]
\begin{centering}
\includegraphics[scale=0.8]{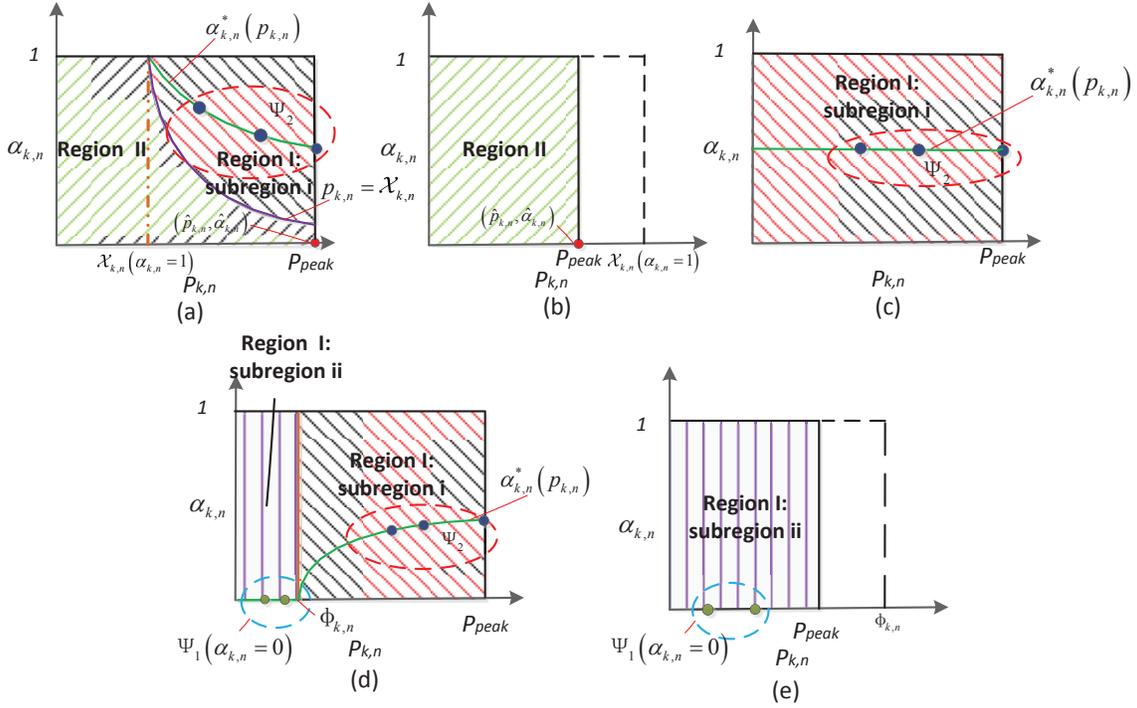}
\vspace{-0.1cm}
 \caption{Five scenarios of feasible regions, where $\phi_{k,n}=\left(\frac{1}{|\beta_{k,n}|^2}-\frac{1}{|h_{k,n}|^2}\right).$}\label{sce1}
\end{centering}
\vspace{-0.3cm}
\end{figure*}

%


\subsubsection{Region II ($0\leq p_{k,n}\leq[\mathcal{X}_{k,n}(\alpha_{k,n})]^+$)}
As we have discussed, $R_{k,n}^s=0$ in this case, which can be true only when $|h_{k,n}|^2<|\beta_{k,n}|^2$. The Lagrangian can thus be rewritten as
\begin{align}
\mathcal{L}'_n
\left(\bs P_n,\bs \alpha_n \right)=
p_{k,n}  \left(\sum_{l \in \mathcal{K}_2}\lambda_l\zeta_l |h_{l,n}|^2- \gamma \right),
\end{align}
which is a linear function of $p_{k,n}$ and is regardless of $\alpha_{k,n}$. Here, we set $\alpha_{k,n}^*=0$ for convenience.
The feasible candidate ($\hat{p}_{k,n}$,$\hat{\alpha}_{k,n}$) in this region can thus be obtained as\footnote{Note that here we assume $\sum_{l \in \mathcal{K}_2}\lambda_l\zeta_l |h_{k,n}|^2> \gamma$. This is because according to the SC allocation policy as we will discuss in later, SC $n$ will be allocated to IR $k$ only if $\mathcal{L}_{n}$ is positive. If $\sum_{l \in \mathcal{K}_2}\lambda_l\zeta_l |h_{k,n}|^2\leq \gamma$, $\mathcal{L}_{n}$ cannot be positive regardless of $p_{k,n}$. As a result, we ignore the case of $\sum_{l \in \mathcal{K}_2}\lambda_l\zeta_l |h_{k,n}|^2\leq \gamma$ without loss of generality.}
\begin{align}
(\hat{p}_{k,n},\hat{\alpha}_{k,n})&=\left(\min\{[\mathcal{X}_{k,n}(\alpha_{k,n}=0)]^+,P_{\rm peak}\},0\right)\nonumber\\
&=(P_{\rm peak},0). \label{eqn:fcr2}
\end{align}
%


It is observed that the feasibility of the above two regions is determined by the channel conditions and the peak power constraint. Five scenarios are illustrated in Fig. \ref{sce1} and explained as follows:

\begin{enumerate}
  \item In \textit{scenario (a)}, $|h_{k,n}|^2<|\beta_{k,n}|^2$ and $P_{{\rm peak}}>\mathcal{X}_{k,n}(\alpha_{k,n}=1)$. Both \textit{Region II} and \textit{subregion i} in \textit{Region I} are feasible.
Note that curve $\alpha_{k,n}^*(p_{k,n})$ and curve $p_{k,n}=\mathcal{X}_{k,n}(\alpha_{k,n})$ intersect at $\left(\mathcal{X}_{k,n}(\alpha_{k,n}=1),1\right)$.
  \item In \textit{scenario (b)}, $|h_{k,n}|^2<|\beta_{k,n}|^2$ and $P_{{\rm peak}}\leq\mathcal{X}_{k,n}(\alpha_{k,n}=1)$. Only \textit{Region II} is feasible.
  \item In \textit{scenario (c)}, $|h_{k,n}|^2=|\beta_{k,n}|^2$. $\alpha_{k,n}^*=\frac{1}{2}$ in this scenario so only \textit{subregion i} in \textit{Region I} is feasible.
\item In \textit{scenario (d)}, $|h_{k,n}|^2>|\beta_{k,n}|^2$ and $P_{\rm peak}>\left(\frac{1}{|\beta_{k,n}|^2}-\frac{1}{|h_{k,n}|^2}\right)$. Two subregions in \textit{Region I} are feasible.
\item In \textit{scenario (e)}, $|h_{k,n}|^2>|\beta_{k,n}|^2$ and $P_{{\rm peak}}\leq\left(\frac{1}{|\beta_{k,n}|^2}-\frac{1}{|h_{k,n}|^2}\right)$. Only \textit{subregion i} in \textit{Region I} is feasible.
\end{enumerate}

Next, we denote $\mathcal{F}$ as the feasible set by combining the above discussions as follows:
\begin{align}
\mathcal{F}=\begin{cases} \Psi_2\cup\{(P_{\rm peak},0)\},&~{\rm if}~P_{{\rm peak}}>\mathcal{X}_{k,n}(\alpha_{k,n}=1)\nonumber\\
&~{\rm and}~|h_{k,n}|^2<|\beta_{k,n}|^2,\nonumber\\
\{(P_{\rm peak},0)\}, &~{\rm if}~P_{{\rm peak}}\leq\mathcal{X}_{k,n}(\alpha_{k,n}=1) \nonumber\\
&~{\rm and}~|h_{k,n}|^2<|\beta_{k,n}|^2,\nonumber\\
\Psi_2,&~{\rm if}~|h_{k,n}|^2=|\beta_{k,n}|^2,\nonumber\\
\Psi_2\cup\Psi_1(\alpha_{k,n}=0),&~{\rm if}~P_{\rm peak}>\left(\frac{1}{|\beta_{k,n}|^2}-\frac{1}{|h_{k,n}|^2}\right)\nonumber\\
&~{\rm and}~|h_{k,n}|^2>|\beta_{k,n}|^2,\nonumber\\
\Psi_1(\alpha_{k,n}=0),&~{\rm if}~P_{\rm peak}\leq\left(\frac{1}{|\beta_{k,n}|^2}-\frac{1}{|h_{k,n}|^2}\right)\nonumber\\
&~{\rm and}~|h_{k,n}|^2>|\beta_{k,n}|^2.
\end{cases}
\end{align}

Given SC $n$ being allocated to IR $k$, the jointly optimized $(p_{k,n}^*,\alpha_{k,n}^*)$ is obtained as
\begin{align}
(p_{k,n}^*,\alpha_{k,n}^*)= \arg \max_{\left(p_{k,n},\alpha_{k,n}\right) \in \mathcal{F}} \mathcal{L}_{n}\left(p_{k,n},\alpha_{k,n}\right).\label{eqn:JO}
\end{align}

The above algorithm is summarized as Algorithm 1.
   \begin{algorithm}[tb]
\caption{Joint optimization of $p_{k,n}^*$ and $\alpha_{k,n}^*$}
\begin{algorithmic}[1]
\IF {$|h_{k,n}|^2>|\beta_{k,n}|^2$}
\IF {$P_{\rm peak}>\left(\frac{1}{|\beta_{k,n}|^2}-\frac{1}{|h_{k,n}|^2}\right)$}
\STATE Compute $\mathcal{F}=\Psi_1(\alpha_{k,n}=0)\cup\Psi_2$ via \eqref{eqn:optp1} and \eqref{eqn:optp2}.
\ELSE
\STATE Compute $\mathcal{F}=\Psi_1(\alpha_{k,n}=0)$ via \eqref{eqn:optp1}.
\ENDIF
\ELSIF {$|h_{k,n}|^2=|\beta_{k,n}|^2$}
\STATE Compute $\mathcal{F}=\Psi_2$ via \eqref{eqn:optp2}.
\ELSIF {$P_{\rm peak}>\mathcal{X}_{k,n}(\alpha_{k,n}=1)$}
\STATE Compute $\mathcal{F}=\Psi_2\cup\{(P_{\rm peak},0)\}$ via \eqref{eqn:optp2}.
\ELSE
\STATE Set $\mathcal{F}=\{(P_{\rm peak},0)\}$.
\ENDIF
\STATE Compute $(p_{k,n}^*,\alpha_{k,n}^*)$ according to \eqref{eqn:JO}.
\end{algorithmic}

\end{algorithm}

%

\subsection{Subcarrier Allocation}

Substituting the optimal $p_{k,n}^*$ and $\alpha_{k,n}^*$ into $\mathcal{L}_n'$, the optimal SC assignment policy is given by
   \begin{eqnarray}
     x_{k,n}^{*}=\begin{cases} ~1, &~{\rm if}~k=k^{*}=\arg \max_{k \in \mathcal{K}_1}\mathcal{L}_{n}'(p_{k,n}^*,\alpha_{k,n}^*)\\
     &~{\rm and}~\max_{k \in \mathcal{K}_1}\mathcal{L}_{n}'(p_{k,n}^*,\alpha_{k,n}^*)>0, \\
     ~0, &~{\rm otherwise}\end{cases}.\label{eqn:opx1}
   \end{eqnarray}
\subsection{Dual Update}

According to \cite{Boyd}, the dual problem is always convex; hence, the subgradient method can be used to update the dual variables to the optimal ones by an iterative procedure:
    \begin{align}
&\lambda_{l}^{t+1}=\left[\lambda_{l}^{t}-\xi_l\left(Q_{l}-\bar{Q}_l\right)\right]^+, \forall l \in \mathcal{K}_2, \label{eqn:updatel}\\
&\gamma^{t+1}=\left[\gamma^{t}-\nu\left( P_{\rm max}-\sum_{n \in \mathcal{N}} \sum_{k \in \mathcal{K}_1} x_{k,n} p_{k,n} \right) \right]^+\label{eqn:updateg},
\end{align}
where $t\geq0$ is the iteration index, $\left[\xi_1,...,\xi_{K_2}\right]$ and $\nu$ are properly designed positive step-sizes.

Note that the subgradient method is an iterative method for solving convex optimization problems in general, and the proposed algorithm is a direct application of the subgradient method to our problem. Thus the convergence and optimality of the proposed algorithm can be guaranteed.

\subsection{Complexity}
The complexity of this iterative algorithm is analyzed as follows.
For each SC, $ \mathcal{O} ({K_2}) $ computations are needed for solving $\Omega_n$ in \eqref{eqn:Omega} and $ \mathcal{O} ({K_1}) $ computations are needed for searching the best IR. Since the optimization is independent at each SC,
the complexity is $\mathcal{O} ({K N})$ for each iteration. Last, the complexity of subgradient based updates is polynomial in the number of dual variables $K_2+1$ \cite{Boyd}.
As a result, the overall complexity of the proposed algorithm for solving problem \eqref{eqn:pbp} is $\mathcal{O} ({(K_2+1)^{q}K N})$, where $q$ is a positive constant. Note that the complexity is polynomial.

Finally, we summarize the overall algorithm for solving problem (13a) in Algorithm 2.
   \begin{algorithm}[tb]
\caption{Optimal Algorithm for Problem \eqref{eqn:pbp}}
\begin{algorithmic}[1]
\REPEAT
\STATE Jointly optimize $p_{k,n}^*$ and $\alpha_{k,n}^*$ for all $k\in \mathcal{K}_1$ and $n \in \mathcal{N}$ according to Algorithm 1.
\STATE Solve SC allocation $x_{k,n}^*$ for all $k \in \mathcal{K}_1$ and $n \in \mathcal{N}$ according to \eqref{eqn:opx1}.
\STATE Update $\bs \lambda$ and $\gamma$ according to \eqref{eqn:updatel} and \eqref{eqn:updateg}, respectively.
\UNTIL {$\bs \lambda$ and $\gamma$ converges.}

\end{algorithmic}

\end{algorithm}
\section{Suboptimal Solution}
The complexity of the optimal algorithm becomes high as $K_1$, $K_2$ and/or $N$ increases, mainly due to the updating of the Lagrange multipliers $\bs \lambda$ and $\gamma$.  
By eliminating the dual updates, in this section, we present an efficient suboptimal algorithm which significantly reduces the complexity.

We design a two-stage algorithm by assuming equal power allocation, i.e., $p_{k,n}=\min\{P_{\rm peak},$ $P_{\rm max}/N\}$, $\forall k \in \mathcal{K}_1, n \in \mathcal{N}$. Here we drop index $k$ and $n$ of $p_{k,n}$ for brevity. In the first stage, for each unsatisfied ER $k$, we select the SC at which ER $k$ has the largest channel gain among all unsatisfied ERs and then assign this SC to the IR $k¡ä$ that
has the largest channel gain among all IRs. The above process is repeated until the minimum
harvested power of all the ERs are satisfied. We denote $N_1$ as the number of SCs assigned in
this stage given in the set $\mathcal N_1$, and $N_2$ as the number of unassigned SCs in the set $\mathcal N_2$.

%
In the second stage, we consider the following problem that is simplified from problem \eqref{eqn:pbp}.
\begin{align}
&\max_{\bs X,\bs \alpha} R_{\rm sum}^s \label{eqn:pb3}\\
{\rm s.t.} ~~~
&\eqref{con:a}-\eqref{con:x2}.\nonumber
\end{align}

Note that all ERs' constraints on required harvested power are removed as they are already achieved after the first stage. The simplified problem \eqref{eqn:pb3} for power allocation, SC assignment and determining transmit power splitting ratios can be regarded as a special case of problem \eqref{eqn:pbp}. Accordingly, we can obtain the optimal transmit power splitting ratios by \eqref{eqn:opta1}. After that, the problem is reduced to a SC assignment problem for weighted sum secrecy rate maximization, which can be optimally solved by a greedy algorithm, i.e., each SC is assigned to the IR having the largest weighted secrecy rate. Note that the ERs can harvest additional energy from the SCs assigned to the IRs in the second stage.

   \begin{algorithm}[tb]
\caption{Suboptimal Algorithm for Problem \eqref{eqn:pbp}}
\begin{algorithmic}[1]
\STATE Set $\mathcal{N}_1=\emptyset$, and $p=\min\{P_{\rm peak},P_{\rm max}/N\}$.
\FOR {Each ER $l$}
\STATE Compute $Q_{l}=\zeta_lp\sum_{n\in \mathcal{N}_1}|h_{l,n}|^2$.
\REPEAT
\STATE Find unassigned SC $n$ that has the largest channel gain for ER $l$.
\STATE Set $\mathcal{N}_1\leftarrow\mathcal{N}_1\cup n$ and assign SC $n$ to IR $k$ having the largest channel gain.
\STATE Determine the optimal transmit power splitting ratio $\alpha_{k,n}^*$ by using \eqref{eqn:opta1}.
\STATE Compute $Q_{l}\leftarrow Q_{l}+\zeta_l p|h_{l,n}|^2$.
\UNTIL{$Q_{l}\geq\bar{Q}_l$}
\ENDFOR
\FOR {The rest $N_2$ of unassigned SCs}
\STATE Determine the optimal transmit power splitting ratios $\alpha_{k,n}^*$ via \eqref{eqn:opta1}, for all $k \in \mathcal{K}_1$ and $n \in \mathcal{N}_2$.
\STATE Solve SC allocation variable $x_{k,n}^*$  for all $k \in \mathcal{K}_1$ and $n \in \mathcal{N}_2$ by using greedy method.
\ENDFOR
\end{algorithmic}
\end{algorithm}


The above suboptimal algorithm is summarized in Algorithm 3. The complexity order of the first stage is $\mathcal{O}(K_2N_1)$ and the complexity order of the second stage is $\mathcal{O}(K_1N_2)$.
Thus the total complexity is thus given as $\mathcal{O}(K_2N_1+K_1N_2)$ which is upper-bounded by $\mathcal{O}(KN)$
and is much lower than that of Algorithm 2.

\section{Numerical Results}
  In this section, we evaluate the performance of the proposed algorithms through extensive simulations. In the simulation setup, a single cell with radius of $200$ meters (m) is considered. The BS is located at the centre of the cell. The carrier frequency is $900$ MHz and the bandwidth is 1 MHz. We assume the noise power $\sigma^2=-83$ dBm, and antenna gains to be 0 dB. The peak transmit power constraint is set to be $P_{\rm peak}=\infty$. We consider $K_1=4$ IRs that are randomly located in the cell with distance to the BS uniformly distributed. For each IR, we set $w_k=1, \forall k \in \mathcal{K}_1$, i.e., we consider the sum secrecy rate of all IRs. We also consider $K_2=4$ ERs that are uniformly distributed within the circle of radius of $2$ m around the BS.\footnote{We consider ERs in general closer to the BS than IRs to receive larger power (versus that of IRs used for decoding information against background noise only). However, under this circumstance, ERs in general have better channel conditions than IRs, and as a result they are more capable of eavesdropping the information sent by the BS \cite{LiuGC13}.} For each ER, we set $\zeta_l=60\%, \forall l \in \mathcal{K}_2$. The channel coefficients consist of both large-scale fading and small-scale fading. The path loss exponent is set to be 3. The small-scale fading is modeled as Rayleigh fading and each channel realization is composed of 8 i.i.d. Rayleigh fading paths. We also assume that all ERs have the same harvested power requirement, i.e., $\bar{Q}_l=\Bar{Q}, \forall l \in \mathcal{K}_2$.

\begin{figure}[t]
\begin{centering}
\includegraphics[scale=.6]{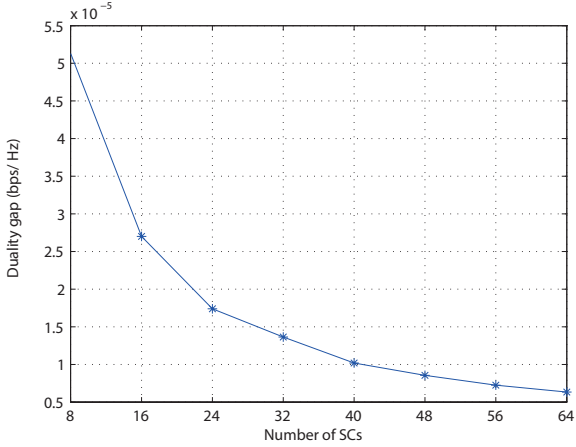}
\vspace{-0.1cm}
 \caption{Duality gap versus number of SCs. }\label{fig:Ngap}
\end{centering}
\vspace{-0.3cm}
\end{figure}
   For performance comparison, we also consider the following benchmarking schemes. First, the fixed transmit power splitting ratio with $\alpha_{k,n}=0.5, \forall k\in \mathcal{K}_1, n\in \mathcal{N}$ is considered for complexity reduction, while the power and SC allocation is still optimized as in Algorithm 2. In this case we drop the index $k$ and $n$ of $\alpha_{k,n}$ for brevity. Second, the SC assignment is fixed (FSA) while the power allocation and transmit power splitting are jointly optimized as in Algorithm 2. Last, we also consider the scheme without using AN (NoAN). It is worth noting that NoAN performs the same as the traditional AN scheme (AN scheme without cancelling) as we have discussed in Remark 1.
First, the duality gaps with different number of SCs $N$ are shown in Fig. \ref{fig:Ngap}. It is observed that duality gap is very small and becomes smaller as $N$ increases. For the case of $N=64$, the gap becomes smaller than $1\times10^{-5}$ bps/Hz, thus is considered to be negligible.
\begin{figure}[t]
\begin{centering}
\includegraphics[scale=.55]{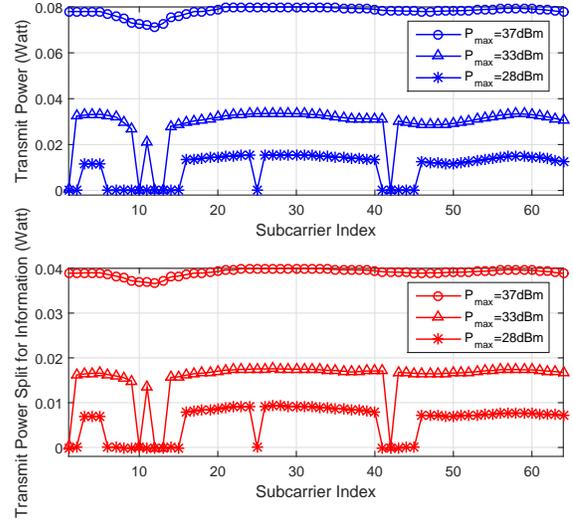}
\vspace{-0.1cm}
 \caption{Transmit power and power split for information source on each SC.}\label{fig:power}
\end{centering}
\vspace{-0.3cm}
\end{figure}

We also show the allocated transmit power and power split for information source over SCs in Fig. \ref{fig:power}, with $\Bar{Q}=100$ $\mu$W. First, we observe that for the case where $P_{\rm max}=37$dBm, the allocated power on each SC is almost uniform, which shows that the suboptimal algorithm that allocates power uniformly over SCs may perform closer to the optimal algorithm as $P_{\rm max}$ increases. In addition, we observe that the power used for information source is approximately one half of the power allocated on each corresponding SC, i.e., the optimal $\alpha_{k,n}\approx0.5$. This is because, according to \eqref{eqn:opta1}, we have the optimal $\alpha_{k,n}^*\approx\frac{1}{2}$ in the high SNR region. In our simulation setup, the noise power is relatively small and results in high $\frac{p_{k,n}}{\sigma^2}$  and thus the optimal  solution $\alpha_{k,n}^*\approx\frac{1}{2}$.

\begin{figure}[t]
\begin{centering}
\includegraphics[scale=.6]{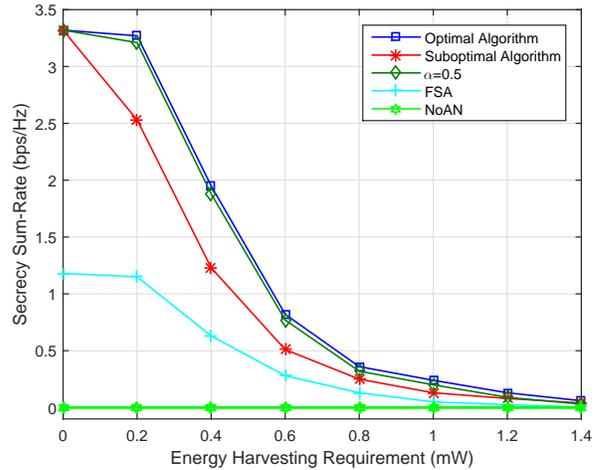}
\vspace{-0.1cm}
 \caption{Achievable secrecy rate $R_{\rm sum}^s$ versus required harvested power $\bar{Q}$. }\label{fig:f1}
\end{centering}
\vspace{-0.3cm}
\end{figure}

\begin{figure}[t]
\begin{centering}
\includegraphics[scale=.6]{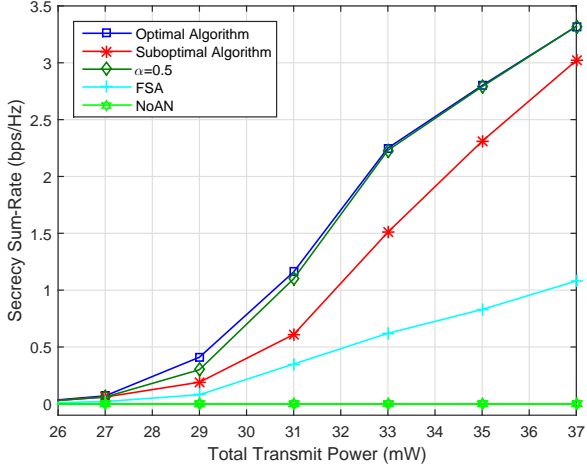}
\vspace{-0.1cm}
 \caption{Achievable secrecy rate $R_{\rm sum}^s$ versus total transmit power constraint $P_{\rm max}$. }\label{fig:f2}
\end{centering}
\vspace{-0.3cm}
\end{figure}

  In Fig. \ref{fig:f1}, the sum secrecy rate $R_{\rm sum}^s$ versus the harvested power requirement $\Bar{Q}$ is shown with $P_{\rm max}=37$ dBm and $N=64$. First, for all schemes (except NoAN), the sum secrecy rate is observed to decrease with increasing $\Bar{Q}$. It is also observed that the suboptimal algorithm and the optimal algorithm outperform FSA and NoAN and the suboptimal algorithm incurs at most $30\%$ loss in secrecy
rate compared to the optimal algorithm.  An interesting observation is that the scheme with $\alpha=0.5$ performs closely to the proposed optimal algorithm, which is in accordance to our previous discussion that $\alpha_{k,n}^*\approx\frac{1}{2}$ in the high SNR region. The poor performance of FSA compared to the proposed algorithms indicates that dynamic SC allocation provides significant gain in terms of sum secrecy rate. Moreover, all considered schemes with AN achieve significant rate-energy gains compared to NoAN, which has almost zero sum secrecy rate even if there is no harvested power requirement. This is because without the effective aid of the AN, the secrecy rate on each SC is positive only when it is assigned to the receiver of largest channel gain\cite{Wang}. However, in our simulation setup, the ERs possess much better channel gains compared to the IRs, due to shorter distances to the BS. As a result, $|h_{k,n}|^2<|\beta_{k,n}|^2$ is almost true for all $n \in \mathcal{N}, k \in \mathcal{K}_1$, and hence no secrecy information can be transmitted at all. This demonstrates the effectiveness of the proposed frequency-domain AN aided approach.

  Fig. \ref{fig:f2} demonstrates the sum secrecy rate $R_{\rm sum}^s$ versus the total transmit power $P_{\rm max}$, with the harvested power constraint set as $\Bar{Q}=100$ $\mu$W and $N=64$. Compared with FSA and NoAN, both proposed optimal and suboptimal algorithms perform better. In addition, it can be observed that suboptimal algorithm performs more closely to the optimal algorithm as the total transmit power increases, which collapses to the observation from Fig. 7 that the allocated power on SCs is more uniformly distributed as transmit power increases. Moreover, the scheme with $\alpha=0.5$ is also observed to perform very closely to the optimal algorithm.

\begin{figure}[t]
\begin{centering}
\includegraphics[scale=.6]{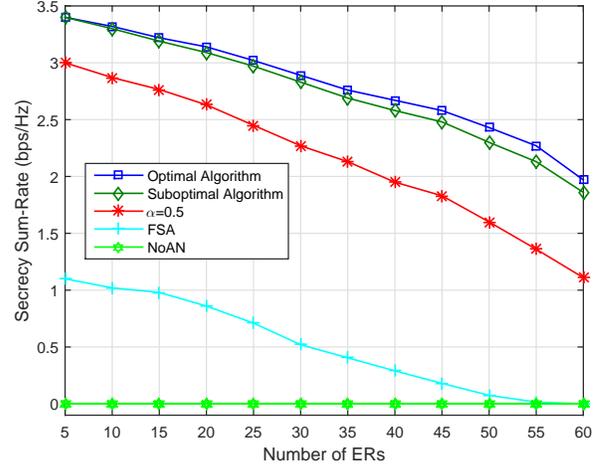}
\vspace{-0.1cm}
 \caption{Achievable secrecy rate $R_{\rm sum}^s$ versus the number of ERs. }\label{fig:f4}
\end{centering}
\vspace{-0.3cm}
\end{figure}

 Fig. \ref{fig:f4} illustrates the sum secrecy rate versus the number of ERs, with the harvested power requirement set as $\Bar{Q}=100$ $\mu$W, $P_{\rm max}=37$ dBm and $N=64$. First, we observe that with the increasing number of ERs, the sum secrecy rate of IRs for all schemes decreases. This is because when a new ER is added in the system, $|\beta_{k,n}|^2$ may increase for all IRs at any SC $n$. As a result, secrecy information is more easily eavesdropped. In addition, with more ERs, more power will be allocated to the SCs for satisfying the requirements of the ERs but not necessarily achieving the maximum sum secrecy rate for IRs. It is also observed that FSA becomes infeasible when the number of ERs is larger than 55, while the proposed algorithms perform with noticeably higher sum secrecy rate.

\section{Conclusion}

    This paper studies the optimal resource allocation for OFDMA-based SWIPT with secrecy constraints. With a proposed frequency-domain AN generation and removal method, we maximize the weighted sum secrecy rate for IRs subject to individual harvested power constraints of ERs by jointly optimizing transmit power and SC allocation as well as transmit power splitting ratios over SCs for AN signals. We proposed an algorithm based on the Lagrange duality to solve the formulated problem with polynomial time complexity. We also proposed a suboptimal algorithm with lower complexity. Through extensive simulations, we showed that the proposed algorithms outperform other heuristically designed schemes with or without using the AN.

\appendices

\section{Proof of Lemma 1}

We consider the following two cases:

\begin{enumerate}
   \item $\alpha_{k,n}\neq0$:
   Equating $r_{k,n}-r_{k,n}^e$ to zero, we obtain

\begin{align}
\frac{|h_{k,n}|^2p_{k,n}}{\sigma^2}&=\frac{|\beta_{k,n}|^2p_{k,n}}{\alpha_{k,n}|\beta_{k,n}|^2p_{k,n}+\sigma^2}.
\end{align}

   We thus have $p_{k,n}=0$ or $p_{k,n}=\mathcal{X}_{k,n}(\alpha_{k,n})$. However, $p_{k,n}$ is always non-negative, so $p_{k,n}=\mathcal{X}_{k,n}(\alpha_{k,n})>0$ can be true only when $|h_{k,n}|^2<|\beta_{k,n}|^2$. Thus, it is shown that $r_{k,n}-r_{k,n}^e=0$ has one root at $p_{k,n}=0$, when $|h_{k,n}|^2\geq|\beta_{k,n}|^2$, and two roots at $p_{k,n}=0$ and $p_{k,n}=\mathcal{X}_{k,n}(\alpha_{k,n})$, when $|h_{k,n}|^2<|\beta_{k,n}|^2$.

For brevity, we define $x\triangleq\alpha_{k,n}$, $y\triangleq p_{k,n}$, $h\triangleq|h_{k,n}|^2/\sigma^2$ and $g\triangleq|\beta_{k,n}|^2/\sigma^2$. When $|h_{k,n}|^2<|\beta_{k,n}|^2\Leftrightarrow h<g$, it follows that
\begin{align}
&\frac{\partial (r_{k,n}-r_{k,n}^e)}{\partial p_{k,n}}
\Bigg{|}_{p_{k,n}=\mathcal{X}_{k,n}(\alpha_{k,n})}
\triangleq \frac{\partial f}{\partial y}\Big{|}_{y=\mathcal{X}_{k,n}(x)}\nonumber\\
=&\frac{g}{\ln2[(1 - g/h)/x - 1]} - \frac{gx}{\ln2[(1 - g/h) - 1]} \nonumber\\
&+ \frac{h(1 - x)}{\ln2((h/g - 1)(x - 1)/x + 1)}\nonumber\\
=&\frac{hx(g - h)(1 - x)}{\ln2(g - h + hx)}\nonumber\\
\geq&0.
\end{align}
Hence, $r_{k,n}-r_{k,n}^e\leq0$ when $0\leq p_{k,n}\leq \mathcal{X}_{k,n}(\alpha_{k,n})$ and $|h_{k,n}|^2<|\beta_{k,n}|^2$, which is equivalent to $0\leq p_{k,n}\leq [\mathcal{X}_{k,n}(\alpha_{k,n})]^+$. On the other hand, $r_{k,n}-r_{k,n}^e>0$ when i) $p_{k,n}> \mathcal{X}_{k,n}(\alpha_{k,n})$ and $|h_{k,n}|^2<|\beta_{k,n}|^2$ or ii) $p_{k,n}> 0$ and $|h_{k,n}|^2\geq |\beta_{k,n}|^2$, which is equivalent to $p_{k,n}>[\mathcal{X}_{k,n}(\alpha_{k,n})]^+$.

   \item $\alpha_{k,n}=0$: In this case, we have
   \begin{equation}\label{easd}
   R_{k,n}^s=\begin{cases}0, &~~{\rm if} ~~ |\beta_{k,n}|^2\geq|h_{k,n}|^2\\
   r_{k,n}-r_{k,n}^e>0, &~~{\rm if} ~~ |\beta_{k,n}|^2<|h_{k,n}|^2\end{cases}.
   \end{equation}
Forcing $[\mathcal{X}_{k,n}]^+\rightarrow+\infty$ when $|\beta_{k,n}|^2\geq|h_{k,n}|^2$ and $[\mathcal{X}_{k,n}]^+=0$ when $|\beta_{k,n}|^2<|h_{k,n}|^2$, \eqref{easd} is equivalently written as
      \begin{equation}
   R_{k,n}^s=\begin{cases}0, &~~{\rm if} ~~ 0\leq p_{k,n}\leq[\mathcal{X}_{k,n}]^+\\
   r_{k,n}-r_{k,n}^e>0, &~~{\rm if} ~~ p_{k,n}>[\mathcal{X}_{k,n}]^+\end{cases}.
   \end{equation}
 \end{enumerate}

Combining the above two cases, we can finally conclude that $R_{k,n}^s=0$ when $0\leq p_{k,n}\leq [\mathcal{X}_{k,n}(\alpha_{k,n})]^+$, while $R_{k,n}^s=r_{k,n}-r_{k,n}^e>0$ when $p_{k,n}>[\mathcal{X}_{k,n}(\alpha_{k,n})]^+$.

The proof is thus completed.
%
%
%
%

\section{Optimal Transmit Power Splitting Ratio for Traditional AN Scheme}
When the AN cannot be cancelled at the intended IR, the secrecy rate in \eqref{r} should be rewritten as
\begin{align}
R_{k,n}^{s,NC}=&[r_{k,n}-r_{k,n}^e]^+\nonumber\\
=&\left[\log_2 \left(1+\frac{(1-\alpha_{k,n})|h_{k,n}|^2p_{k,n}}{\alpha_{k,n}|h_{k,n}|^2p_{k,n}+\sigma^2}\right)\right.\nonumber\\
&\left.-\log_2 \left(1+\frac{(1-\alpha_{k,n})|\beta_{k,n}|^2p_{k,n}}{\alpha_{k,n}|\beta_{k,n}|^2p_{k,n}+\sigma^2}\right)\right]^+.
\end{align}

We first consider the problem $\max_{\alpha_{k,n}} R_{k,n}^{s,NC}$ by focusing on the following two cases:
\begin{enumerate}
  \item For the case that $|h_{k,n}|^2>|\beta_{k,n}|^2$, we have $R_{k,n}^{s,NC}>0$ and
  \begin{align}
  &\frac{\partial R_{k,n}^{s,NC}}{\partial \alpha_{k,n}}\nonumber\\
  =&-\frac{1}{\ln 2}\frac{(|h_{k,n}|^2-|\beta_{k,n}|^2)\sigma^2p_{k,n}}{(\alpha_{k,n}|h_{k,n}|^2p_{k,n}+\sigma^2)(\alpha_{k,n}|\beta_{k,n}|^2p_{k,n}+\sigma^2)}\nonumber\\
  \leq&0.
  \end{align}

  Thus, we have that $R_{k,n}^{s,NC}$ is monotonically non-increasing with respect to $\alpha_{k,n}$ and the optimal solution is given by $\alpha_{k,n}^*=0, \forall k,n$.
  \item For the case that $|h_{k,n}|^2\leq|\beta_{k,n}|^2$, we have $R_{k,n}^{s,NC}=0$ regardless of $\alpha_{k,n}$.
\end{enumerate}

Combining the above two cases, we conclude that $\alpha_{k,n}^*=0, \forall k,n$, is always optimal to maximize the secrecy rate using traditional AN scheme without cancelation at the receiver, i.e., the traditional AN scheme performs no better than the transmission without AN.

In addition, to show $\alpha_{k,n}^*=0, \forall k,n$ is also the optimal solution to the sum secrecy rate maximization problem under energy harvesting constraints similar to problem (13a), we consider the following problem
\begin{align}
 &\max_{\bs \alpha} \sum_{k\in \mathcal{K}_1}\sum_{n\in\mathcal{N}} R_{k,n}^{s,NC}\label{eqn:pbff}\\
&{\rm s.t.}~~~ (13{\rm b})-(13{\rm g})\nonumber.
\end{align}

We can show the decomposed Lagrangian on each SC $\mathcal{L}_n$ of problem \eqref{eqn:pbff} is obtained in \eqref{subl} by replacing $R_{k,n}^{s}$ with $R_{k,n}^{s,NC}$ and
\begin{align}
\frac{\partial\mathcal{L}_n}{\partial \alpha_{k,n}}=w_k\frac{\partial R_{k,n}^{s,NC}}{\partial \alpha_{k,n}}\leq0.
\end{align}

Thus, the solution $\alpha_{k,n}^*=0, \forall k,n$ also holds optimality for problem \eqref{eqn:pbff}.

\section{Proof of Lemma 2}
By applying the KKT (Karush-Kuhn-Tucker) conditions \cite{Boyd}, we obtain
\begin{align}
\alpha_{k,n}^*(p_{k,n})=\left[\frac{1}{2}+\frac{\sigma^2}{2p_{k,n}}\left(\frac{1}{|h_{k,n}|^2}-\frac{1}{|\beta_{k,n}|^2}\right)\right]_0^1,
\end{align}
for all $k \in \mathcal{K}_1, n \in \mathcal{N}$, where $[\cdot]_a^b\triangleq\min\{\max\{\cdot,a\},b\}.$

When $|h_{k,n}|^2<|\beta_{k,n}|^2$, $p_{k,n}>[\mathcal{X}_{k,n}(\alpha_{k,n})]^+=\mathcal{X}_{k,n}(\alpha_{k,n})$, we thus have
\begin{align}
  \alpha_{k,n}^*&=\left[\frac{1}{2}+\frac{(|\beta_{k,n}|^2-|h_{k,n}|^2)\sigma^2}{2|\beta_{k,n}|^2|h_{k,n}|^2p_{k,n}}\right]_0^1\nonumber\\
                &<\frac{1}{2}+\frac{(|\beta_{k,n}|^2-|h_{k,n}|^2)\sigma^2}{2|\beta_{k,n}|^2|h_{k,n}|^2\mathcal{X}_{k,n}(\alpha_{k,n})}\nonumber\\
                &=\frac{1}{2}+\frac{\alpha_{k,n}^*}{2}\nonumber\\
                &<1.
  \end{align}

When $|h_{k,n}|^2\geq|\beta_{k,n}|^2$, $p_{k,n}>[\mathcal{X}_{k,n}(\alpha_{k,n})]^+=0$, we thus have
\begin{align}
\alpha_{k,n}^*=\left[\frac{1}{2}+\frac{\sigma^2}{2p_{k,n}}\left(\frac{1}{|h_{k,n}|^2}-\frac{1}{|\beta_{k,n}|^2}\right)\right]_0^1<\frac{1}{2}.
\end{align}

To conclude the above two cases, we have $\alpha_{k,n}^*<1$ is always true for $p_{k,n}\geq [\mathcal{X}_{k,n}(\alpha_{k,n})]^+$.
Thus the optimal $\alpha_{k,n}^*$ with given $p_{k,n}$ is rewritten as
 \begin{align}
\alpha_{k,n}^*(p_{k,n})=\left[\frac{1}{2}+\frac{\sigma^2}{2p_{k,n}}\left(\frac{1}{|h_{k,n}|^2}-\frac{1}{|\beta_{k,n}|^2}\right)\right]^+,
\end{align}
for all $k \in \mathcal{K}_1, n \in \mathcal{N}$.

The proof is thus completed.

\bibliographystyle{IEEEtran}
\bibliography{IEEEabrv,TW-Dec-14-1842}

\end{document}